\documentclass[preprint,superscriptaddress,aps]{revtex4-1}
\usepackage[utf8]{inputenc}
\usepackage{amsmath}
\usepackage{graphicx}

\makeatletter

\DeclareTextSymbolDefault{\textquotedbl}{T1}

\usepackage{amsfonts}
\usepackage{slashed}

\global\long\def\sl#1{\slashed{#1}}

\makeatother

\begin{document}
\title{Perturbative generation of photon Lorentz violating terms from a pseudo-tensor
Lorentz-breaking extension of QED}
\author{A. J. G. Carvalho}
\affiliation{Departamento de Física, Universidade Federal da Paraíba,
 Caixa Postal 5008, 58051-970, João Pessoa, Paraíba, Brazil}
 \affiliation{ Departamento de Ciências Naturais, Universidade do Estado do Pará-UEPA, 
		68502-100, Marabá, Pará, Brasil}
\email{gomescarvalhoantoniojose, andrem.physics@gmail.com, jroberto, petrov@fisica.ufpb.br}

\author{A. F. Ferrari}
\affiliation{Universidade Federal do ABC, Centro de Ciências Naturais e Humanas,
Rua Santa Adélia, 166, 09210-170, Santo André, SP, Brasil}
\email{alysson.ferrari@ufabc.edu.br}

\affiliation{Indiana University Center for Spacetime Symmetries, Indiana University,
Bloomington, Indiana 47405-7105, USA}
\author{A. M. de Lima}
\affiliation{Departamento de Física, Universidade Federal da Paraíba~~~\\
 Caixa Postal 5008, 58051-970, João Pessoa, Paraíba, Brazil}
\email{gomescarvalhoantoniojose, andrem.physics@gmail.com, jroberto, petrov@fisica.ufpb.br}

\author{J. R. Nascimento}
\affiliation{Departamento de Física, Universidade Federal da Paraíba~~~\\
 Caixa Postal 5008, 58051-970, João Pessoa, Paraíba, Brazil}
\email{gomescarvalhoantoniojose, andrem.physics@gmail.com, jroberto, petrov@fisica.ufpb.br}

\author{A. Yu. Petrov}
\affiliation{Departamento de Física, Universidade Federal da Paraíba~~~\\
 Caixa Postal 5008, 58051-970, João Pessoa, Paraíba, Brazil}
\email{gomescarvalhoantoniojose, andrem.physics@gmail.com, jroberto, petrov@fisica.ufpb.br}

\begin{abstract}
We consider an extended QED with the addition of a dimension-five
Lorentz-breaking coupling between spinor and gauge fields, involving
a pseudo-tensor $\kappa^{\mu\nu\lambda\rho}$. The specific form of
the Lorentz violating coupling considered by us have been suggested
in other works, and some of its consequences at the classical level
were already studied. Here, we investigate the consequences of this
specific form of Lorentz violation at the quantum level, evaluating
the one loop corrections to the gauge field two-point function, both
at zero and at finite temperature. We relate the terms that are generated
by quantum corrections with the photon sector of the Standard Model
Extension, discussing the possibility of establishing experimental
bounds on $k^{\mu\nu\rho\sigma}$. From the dispersion relations in
the resulting theory, we discuss its consistency from the causality
viewpoint.
\end{abstract}
\maketitle

\section{Introduction}

One of the most important directions of research related to Lorentz
symmetry violations is the study of the impacts of different Lorentz-breaking
extensions of the known field theory models at the classical and quantum
levels. The concept of effective field theories is very appropriate
for this end, since it allows the incorporation in the Standard Model,
which is known to describe accurately physics in the current accessible
energies, of small Lorentz violating terms, assumedly originated in
some more fundamental theory at very small length scales. Originally,
a Lorentz violating (LV) extension of the Standard Model, called the
Standard Model Extension (SME), including minimal (mass dimension
up to four) LV terms, was been presented in \cite{Kostelecky}, providing
a systematic framework for theoretical and experimental investigations.
From the theoretical standpoint, once LV terms are included in the
Lagrangian, it is an interesting question how much of the known structure
of Quantum Field Theories is preserved, for instance, whether consistent
quantum corrections can be calculated. Regarding the perturbative
generation of LV terms in one sector of the theory (e.g. the photon
sector) originating from another sector, such as the coupling between
the photon and a fermion, some known examples are the axial coupling
used to generate the CFJ term \cite{CFJ,JackiwKostelecky} and the
magnetic one used to generate the aether term \cite{aether}. Nevertheless,
other couplings deserve to be studied as well, and in particular a
systematic study of higher dimensional operators, called non minimal
LV terms, have been worked out in recent years. Some interesting results
for other couplings have been obtained in \cite{Casana3,BorgesHDQED,Belich1}
where quantum corrections in the extended QED with dimension-five
tensor couplings have been considered, as well as\,\cite{BorgesAxion}
where a possible generation of an axion-photon coupling from a LV
model was discussed. General discussion of higher-dimensional LV operators
can be found in\,\cite{Anton,nonMinimalLV1,nonMInimalLV2}, and in
\cite{Ferrnew} a recent study of some specific dimension-six operators
was reported.

An interesting example of a non minimal interaction based on a constant
pseudo tensor $\kappa^{\mu\nu\lambda\rho}$ has been introduced in\,\cite{AraujoAxial,AraujoGeneralCoupling},
in the form $\kappa^{\mu\nu\lambda\rho}\bar{\psi}\sigma_{\mu\nu}\gamma_{5}\psi F_{\lambda\rho}$,
where $\sigma_{\mu\nu}=\frac{i}{2}\left[\gamma_{\mu},\gamma_{\gamma}\right]$.
In those works, the contribution from this LV coupling to the Dirac
equation was worked out in detail, and some experimental constrains
were obtained. A natural problem is the study of the perturbative
implications of this coupling, since, in principle, its contributions
to the photon sector could allow us to use the stringent experimental
constraints on photon physics to put even stronger limits on $\kappa^{\mu\nu\lambda\rho}$.
Besides that, theories with non-minimal LV are potentially plagued
by problems regarding theoretical consistency, such as violations
of causality and/or unitarity, and we will also investigate some of
these questions in our model.

In this paper, we consider one-loop corrections in the extended spinor
QED involving this pseudotensor LV coupling. We explicitly demonstrate
that the consistent treatment of these interactions will require introduction
of specific LV terms in the purely gauge sector, already at tree level.
Afterwards, we discuss plane wave solutions and dispersion relations
in the resulting LV extension of Maxwell electrodynamics, giving some
estimations for the Lorentz-breaking coefficients based on known experimental
data, as well as discussing the possibility of non-causal wave propagation.

The structure of the paper is as follows. In Section 2, we carry out
the one-loop calculations of the two-point functions of the gauge
and spinor fields, for zero and finite temperature. In Section 3,
we discuss the structure of the minimal LV coefficients in the gauge
sector in the resulting extension of QED, and the general structure
of dispersion relations in this theory is considered in Section 4,
allowing us to discuss the causality of wave propagation. In Section
5, we consider the non minimal LV contributions, again discussing
the plane wave solutions and related causality issues. Finally, in
the section 6 we present our conclusions.

\section{\label{sec:Perturbative-Calculations}Definition of the Model and
Perturbative Calculations}

We start with a QED-like Lagrangian including a pseudo tensor Lorentz
violating coupling, given by 
\begin{equation}
\mathcal{L}_{0}=\bar{\psi}\left(i\sl{\partial}-m-e\sl A-ig\kappa^{\mu\nu\lambda\rho}\sigma_{\mu\nu}\gamma_{5}F_{\lambda\rho}\right)\psi\thinspace.\label{eq:defModel}
\end{equation}
The LV pseudotensor satisfies $\kappa^{\mu\nu\lambda\rho}=-\kappa^{\nu\mu\lambda\rho}$
and $\kappa^{\mu\nu\lambda\rho}=-\kappa^{\mu\nu\rho\lambda}$, but
for simplicity we assume also that $\kappa^{\mu\nu\lambda\rho}=\kappa^{\lambda\rho\mu\nu}$,
so that it has essentially the same symmetry properties as the CPT
even $\kappa_{F}$ coefficient present in the photon sector of the
SME\,\cite{LVQED}. We also remark that $\kappa^{\mu\nu\lambda\rho}\bar{\psi}\sigma_{\mu\nu}\gamma_{5}\psi F_{\lambda\rho}$
is a dimension five operator, which therefore is embedded into the
non minimal sector of the SME. The LV coefficient $\kappa^{\mu\nu\lambda\rho}$
is chosen to be dimensionless, whereas the coupling constant $g$
has dimensions of inverse of mass.

Before proceeding with the perturbative calculations, some comments
are in order.First, a Lagrangian similar to Eq.\,\eqref{eq:defModel},
without the $\gamma_{5}$, was already extensively studied in the
literature (see f.e.\,\cite{Casana3,Belich1}). The particular non-minimal
LV coupling in Eq.\,\eqref{eq:defModel} have been first introduced
in\,\cite{AraujoAxial,AraujoGeneralCoupling}, where some of its
tree-level consequences have been discussed. The presence of the $\gamma_{5}$
in Eq.\,\eqref{eq:defModel} inverts the parity properties of the
corresponding operator, and because of that, the LV coefficient $\kappa^{\mu\nu\lambda\rho}$
in our case is a pseudo tensor, instead of being a tensor as in\,\cite{Casana3,Belich1}.
Despite that, we are still dealing with a CPT even Lorentz violating
operator. That may be seen by relating the non minimal LV coupling
in Eq.\,\eqref{eq:defModel} with the general parametrization of
the non minimal QED extension proposed in\,\cite{Ding} by using
the identity $\sigma^{\mu\nu}\gamma^{5}=\frac{i}{2}\epsilon^{\mu\nu\alpha\beta}\sigma_{\alpha\beta}$,
thus obtaining 
\begin{equation}
\mathcal{L}_{0}\supset-\frac{1}{4}H_{F}^{\left(5\right)\mu\nu\alpha\beta}F_{\alpha\beta}\bar{\psi}\sigma_{\mu\nu}\psi\thinspace,
\end{equation}
where 
\begin{equation}
H_{F}^{\left(5\right)\mu\nu\alpha\beta}=-2g\kappa^{\mu\nu\lambda\rho}\epsilon_{\lambda\rho}^{\hphantom{\lambda\rho}\alpha\beta}\thinspace.\label{eq:H5correspondence}
\end{equation}
Therefore, the LV coupling considered by us can be seen as a particular
form of the general CPT even coefficient $H_{F}^{\left(5\right)\mu\nu\alpha\beta}$
introduced in\,\cite{Ding}. Very few experimental constraints on
these LV coefficients exist in the literature. Some constraints for
electrons and protons were first reported on\,\cite{Ding}, and quite
recently improved by two to three orders of magnitude as reported
in\,\cite{BASE1,BASE2}, in experiments involving trapped antiprotons
at CERN. The best experimental result is that some components of the
dimensional combination $H_{F}^{\left(5\right)\mu\nu\alpha\beta}\sim g\kappa^{\mu\nu\lambda\rho}\epsilon_{\lambda\rho}^{\hphantom{\lambda\rho}\alpha\beta}$
have constraints of order $10^{-8}\,\text{GeV}^{-1}$ for protons,
or $10^{-10}\,\text{GeV}^{-1}$ for electrons. One of our results
will be that, via radiative corrections, the specific $H_{F}^{\left(5\right)}$
coefficient we consider will induce birefringent effects in the photon
sector, thus opening a possible window for more stringent constraints
derived from this sector.

\begin{figure}
\begin{centering}
\includegraphics[width=0.3\paperwidth]{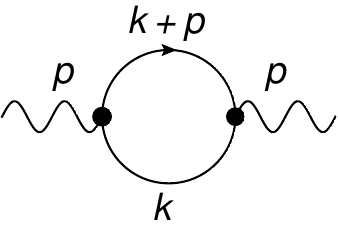}\,\,\includegraphics[width=0.3\paperwidth]{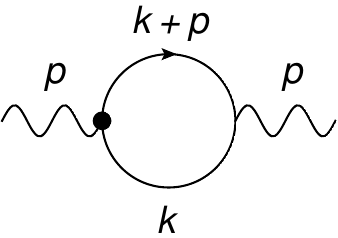}
\par\end{centering}
\caption{\label{fig:Loops}One-loop LV corrections to the photon propagator;
the LV vertex $ig\kappa^{\mu\nu\lambda\rho}\sigma_{\mu\nu}\gamma_{5}$
is represented by the black dot, while straight and wavy lines represent
the free fermion and photon propagators, respectively.}
\end{figure}

From Eq.\,\eqref{eq:defModel}, one obtains the Feynman rules that
can be used do calculate perturbative corrections to the photon two-point
function. Besides the usual free fermion propagator, 
\begin{equation}
S\left(p\right)=\frac{\sl p+m}{p^{2}-m^{2}}\thinspace,
\end{equation}
and the usual QED vertex $ie\gamma^{\mu}$, we also have the LV trilinear
vertex proportional to $ig\kappa^{\mu\nu\lambda\rho}\sigma_{\mu\nu}\gamma_{5}$,
which we will distinguish from the LI vertex with a black dot in the
diagrams. In Fig.\,\eqref{fig:Loops} we present the two LV corrections
to the photon propagator that arise at one loop, which will be calculated
in this section.

We start by the diagram with two LV vertices, the left one in Fig.\,\ref{fig:Loops},
which corresponds to the following expression, 
\begin{equation}
S_{\kappa\kappa}\left(p\right)=\mathrm{tr}\int \frac{d^4k}{(2\pi)^4}\left(ig\kappa^{\rho\sigma\mu\nu}\sigma_{\rho\sigma}\gamma_{5}\right)i\frac{\sl k+\sl p+m}{\left(k+p\right)^{2}-m^{2}}\left(ig\kappa^{\lambda\epsilon\alpha\beta}\sigma_{\lambda\epsilon}\gamma_{5}\right)i\frac{\sl k+m}{k^{2}-m^{2}}\thinspace F_{\mu\nu}\left(p\right)F_{\alpha\beta}\left(p\right)\thinspace.\label{eq:1}
\end{equation}
It will be convenient to introduce the notation 
\begin{equation}
\bar{F}_{\mu\nu}=\kappa_{\mu\nu\alpha\beta}F^{\alpha\beta}\thinspace,\label{eq:Fbar}
\end{equation}
for the contraction of the electromagnetic field strength with the
LV tensor. After some manipulations we write 
\begin{align}
S_{\kappa\kappa}\left(p\right)= & g^{2}\bar{F}^{\rho\sigma}\left(p\right)\bar{F}^{\lambda\epsilon}\left(p\right)\int \frac{d^4k}{(2\pi)^4}\frac{1}{\left(\left[k+p\right]^{2}-m^{2}\right)\left(k^{2}-m^{2}\right)}\times\nonumber \\
 & \times\mathrm{tr}\left(\sigma_{\rho\sigma}\gamma_{5}\sl k\sigma_{\lambda\epsilon}\gamma_{5}\sl k+\sigma_{\rho\sigma}\gamma_{5}\sl p\sigma_{\lambda\epsilon}\gamma_{5}\sl k+m^{2}\sigma_{\rho\sigma}\gamma_{5}\sigma_{\lambda\epsilon}\gamma_{5}\right)\thinspace.\label{eq:2}
\end{align}
Using that $k_{\mu}k_{\nu}=\frac{1}{4}k^{2}\eta_{\mu\nu}$ under integration,
we find that the first term in the equation above, at $p=0$, vanishes since $\gamma^{\alpha}\sigma^{\mu\nu}\gamma_{\alpha}=0$.
Calculating the relevant traces, from the second term inside the brackets
in Eq.\,\eqref{eq:2}, we have 
\begin{equation}
S_{\kappa\kappa}^{\left(a\right)}\left(p\right)=4g^{2}\bar{F}^{\sigma\rho}\bar{F}^{\lambda\epsilon}\int \frac{d^4k}{(2\pi)^4}\frac{k_{\lambda}\left(p_{\sigma}\eta_{\rho\epsilon}-p_{\rho}\eta_{\sigma\epsilon}\right)+k_{\epsilon}\left(p_{\rho}\eta_{\sigma\lambda}-p_{\sigma}\eta_{\rho\lambda}\right)-p^{\mu}k^{\nu}\varepsilon_{\mu\rho\sigma}^{\phantom{\mu\rho\sigma}\alpha}\varepsilon_{\nu\lambda\epsilon\alpha}}{\left(\left[k+p\right]^{2}-m^{2}\right)\left(k^{2}-m^{2}\right)}\thinspace,
\end{equation}
which, after introduction of a proper Feynman parameter, and with adding the analogous cointribution arising from the first term in \eqref{eq:2} at $p\neq 0$, can be cast as 
\begin{align}
S_{\kappa\kappa}^{\left(a\right)}\left(p\right)= & -4g^{2}\bar{F}^{\sigma\rho}\bar{F}^{\lambda\epsilon}\times\\
 & \times\int_{0}^{1}dx x(1-x)\thinspace\int \frac{d^4k}{(2\pi)^4}\frac{p_{\lambda}\left(p_{\sigma}\eta_{\rho\epsilon}-p_{\rho}\eta_{\sigma\epsilon}\right)+p_{\epsilon}\left(p_{\rho}\eta_{\sigma\lambda}-p_{\sigma}\eta_{\rho\lambda}\right)-p^{\mu}p^{\nu}\varepsilon_{\mu\rho\sigma}^{\phantom{\mu\rho\sigma}\alpha}\varepsilon_{\nu\lambda\epsilon\alpha}}{\left(k^{2}+x\left[1-x\right]p^{2}-m^{2}\right)^{2}}\thinspace.\nonumber 
\end{align}
From the third and last term inside the brackets in Eq.\,\eqref{eq:2},
after similar manipulations we arrive at 
\[
S_{\kappa\kappa}^{\left(b\right)}\left(p\right)=m^{2}g^{2}\left(\bar{F}^{\sigma\rho}\bar{F}_{\sigma\rho}-\bar{F}^{\sigma\rho}\bar{F}_{\rho\sigma}\right)\int_{0}^{1}dx\int \frac{d^4k}{(2\pi)^4}\frac{1}{\left(k^{2}+x\left[1-x\right]p^{2}-m^{2}\right)^{2}}\thinspace.
\]
We notice that for both contributions the momentum integral to be
calculated is 
\begin{eqnarray}
I&=&\mu^{\varepsilon}\int\frac{d^{4-\varepsilon}k}{\left(2\pi\right)^{4-\varepsilon}}\frac{1}{\left(k^{2}+x\left[1-x\right]p^{2}-m^{2}\right)^{2}}=\nonumber\\
&=&\frac{i}{\left(4\pi\right)^{2}}\left(\frac{2}{\varepsilon}-\gamma-\ln\left[\frac{x\left(x-1\right)p^{2}+m^{2}}{\mu^2}\right]+\ln (4\pi)\right)\thinspace,\label{int}
\end{eqnarray}
where dimensional regularization has been used. For convenience, in what follows we redefine $4\pi\mu^2\to \mu^2$. The $x$ integrals
can be calculated exactly, and using the symmetry properties of the
tensor $\kappa_{\mu\nu\alpha\beta}$, we obtain the one-loop, second
order in LV contribution to the photon propagator, 
\begin{eqnarray}
S_{\kappa\kappa}\left(p\right) & = & \frac{i}{\left(4\pi\right)^{2}}\left\{ -4g^{2}\left[4\bar{F}_{\phantom{\sigma}\epsilon}^{\sigma}\bar{F}^{\lambda\epsilon}p_{\lambda}p_{\sigma}-\bar{F}^{\sigma\rho}\bar{F}^{\lambda\epsilon}p^{\mu}p^{\nu}\varepsilon_{\mu\rho\sigma}^{\phantom{\mu\rho\sigma}\alpha}\varepsilon_{\nu\lambda\epsilon\alpha}\right]\right.\times\\
 & \times &
 \Big[\frac{1}{6}\big(\frac{2}{\varepsilon}-\gamma-\ln\frac{m^2}{\mu^2}\big)
-\frac{5}{18}+\sqrt{\frac{4m^2-p^2}{p^2}}\arctan\sqrt{\frac{p^2}{4m^2-p^2}}-
\frac{4m^2-p^2}{6p^2}+\nonumber\\&+& \frac{1}{6}(\frac{4m^2-p^2}{p^2})^{3/2}\arctan\sqrt{\frac{p^2}{4m^2-p^2}}
\Big]
\nonumber \\
 & + & \left.2m^{2}g^{2}\bar{F}^{\sigma\rho}\bar{F}_{\sigma\rho}\left[\frac{2}{\varepsilon}-\gamma-\ln\left(\frac{m^{2}}{\mu^{2}}\right)-\frac{2\sqrt{4m^{2}-p^{2}}}{p}\tan^{-1}\left(\frac{p}{\sqrt{4m^{2}-p^{2}}}\right)-2\right]\right\} \thinspace.\nonumber 
\end{eqnarray}
We note that this contribution, being of the second order in the Lorentz-breaking
parameters $\kappa^{\mu\nu\lambda\rho}$, is extremely small and expected
to be subdominant. Nevertheless, we will study some of its possible
consequences which can be relevant when higher-order contributions
are discussed.

Now we calculate the diagram with one LV vertex, the right one in
Fig.\,\ref{fig:Loops}, which provides us with 
\begin{align}
S_{\kappa}\left(p\right) & =\mathrm{tr}\int\frac{d^4k}{(2\pi)^4}ig\kappa^{\mu\nu\lambda\rho}\sigma_{\mu\nu}\gamma_{5}i\frac{\sl k+\sl p+m}{\left(k+p\right)^{2}-m^{2}}ie\gamma^{\alpha}i\frac{\sl k+m}{k^{2}-m^{2}}F_{\lambda\rho}A_{\alpha}\nonumber \\
 & =4imeg\bar{F}^{\mu\nu}A^{\alpha}\int\frac{d^4k}{(2\pi)^4}\frac{\varepsilon_{\mu\nu\alpha\beta}p^{\beta}}{\left(\left[k+p\right]^{2}-m^{2}\right)\left(k^{2}-m^{2}\right)}\thinspace,
\end{align}
where we have used that ${\rm tr}\left(\gamma_{5}\gamma_{\mu}\gamma_{\nu}\gamma_{\alpha}\gamma_{\beta}\right)=-4i\varepsilon_{\mu\nu\alpha\beta}$.
We can rewrite the above expression by noticing that 
\begin{eqnarray}
\varepsilon_{\mu\nu\alpha\beta}A^{\alpha}p^{\beta} & = & \ \frac{i}{2}\varepsilon_{\mu\nu\alpha\beta}F^{\alpha\beta}=i\tilde{F}_{\mu\nu},
\end{eqnarray}
where we have defined the dual electromagnetic field tensor as 
\begin{equation}
\tilde{F}_{\mu\nu}=\frac{1}{2}\varepsilon_{\mu\nu\alpha\beta}F^{\alpha\beta}\thinspace,\label{eq:dualF}
\end{equation}
thus obtaining 
\[
S_{\kappa}\left(p\right)=-4meg\bar{F}^{\mu\nu}\tilde{F}_{\mu\nu}\int\frac{d^4k}{(2\pi)^4}\frac{1}{\left(\left[k+p\right]^{2}-m^{2}\right)\left(k^{2}-m^{2}\right)}\thinspace.
\]
The remaining momentum integral has already been calculated, resulting
in 
\begin{equation}
S_{\kappa}\left(p\right)=-\frac{4imeg\bar{F}^{\mu\nu}\tilde{F}_{\mu\nu}}{\left(4\pi\right)^{2}}\left(\frac{2}{\varepsilon}-\gamma-\ln\left(\frac{m^{2}}{\mu^{2}}\right)+\frac{2\sqrt{4m^{2}-p^{2}}}{p}\tan^{-1}\left[\frac{p}{\sqrt{4m^{2}-p^{2}}}\right]-2\right).
\end{equation}

\begin{figure}
\begin{centering}
\includegraphics{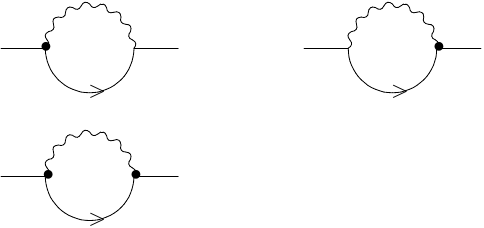}
\par\end{centering}
\caption{\label{fig2}One loop LV contributions to the fermion propagator.}
\end{figure}

One can also calculate the corrections induced by the LV insertions
in the one-loop corrections to the fermion propagator, by considering
the diagrams depicted in Fig.\,\ref{fig2}. The corresponding expressions
are given by
\begin{equation}
\Sigma_{1}(p)=eg\kappa^{\mu\nu\lambda\rho}\bar{\psi}(-p)\sigma_{\mu\nu}\gamma_{5}\int\frac{d^{4}k}{(2\pi)^{4}}\frac{\slashed k-m}{k^{2}-m^{2}}\gamma^{\alpha}\frac{1}{(p-k)^{2}}(p_{\lambda}-k_{\lambda})\eta_{\rho\alpha}\psi(p)\thinspace,
\end{equation}

\begin{equation}
\Sigma_{2}(p)=-eg\kappa^{\mu\nu\lambda\rho}\bar{\psi}(-p)\int\frac{d^{4}k}{(2\pi)^{4}}\gamma^{\alpha}\frac{\slashed k-m}{k^{2}-m^{2}}\frac{1}{(p-k)^{2}}\sigma_{\mu\nu}\gamma_{5}(p_{\lambda}-k_{\lambda})\eta_{\rho\alpha}\psi(p)\thinspace,
\end{equation}

\begin{eqnarray}
\Sigma_{3}(p)&=&-2g^{2}\kappa^{\mu\nu\lambda\rho}\kappa^{\mu'\nu'\lambda'\rho}\bar{\psi}(-p)\int\frac{d^{4}k}{(2\pi)^{4}}\sigma_{\mu\nu}\gamma_{5}\frac{\slashed k-m}{k^{2}-m^{2}}\frac{1}{(p-k)^{2}}\sigma_{\mu'\nu}\gamma_{5}\psi(p)\times\nonumber\\&\times& (p_{\lambda}-k_{\lambda})(p_{\lambda'}-k_{\lambda'})\eta_{\rho\rho'}\thinspace,
\end{eqnarray}
which, after integration, lead to

\begin{equation}
\Sigma_{1}(p)=\frac{eg\kappa^{\mu\nu\lambda\rho}}{16\pi^{2}\epsilon}\psi(-p)\sigma_{\mu\nu}\gamma_{5}\left(\left(\frac{m^{2}}{4}-\frac{p^{2}}{12}\right)\gamma_{\lambda}\gamma_{\rho}+\left(\frac{\slashed p}{6}-\frac{m}{2}\right)p_{\lambda}\gamma_{\rho}\right)\psi(p)+\cdots\thinspace,
\end{equation}

\begin{equation}
\Sigma_{2}(p)=\frac{eg\kappa^{\mu\nu\lambda\rho}}{16\pi^{2}\epsilon}\psi(-p)\left(\left(\frac{m^{2}}{4}-\frac{p^{2}}{12}\right)\gamma_{\lambda}\gamma_{\rho}-\left(\frac{\slashed p}{6}-\frac{m}{2}\right)p_{\lambda}\gamma_{\rho}\right)\sigma^{\mu\nu}\gamma_{5}\psi(p)+\cdots\thinspace,
\end{equation}

\begin{eqnarray}
\Sigma_{3}(p) & =&\frac{2g^{2}}{16\pi^{2}\epsilon}\kappa^{\mu\nu\lambda\rho}\kappa^{\mu'\nu'\lambda'\rho}\eta_{\rho\rho'}\sigma_{\mu\nu}\bar{\psi}(-p)\times\nonumber \\
&\times & \int_{0}^{1}dx\left((m+\slashed px)p_{\lambda}p_{\lambda'}(1-x)^{2}-\frac{1}{4}(m+\slashed px)\eta_{\lambda\lambda'}[m^{2}(1-x)-p^{2}x(1-x)]-\right.\nonumber \\
 & -&\left.\frac{1}{4}(1-x)[m^{2}(1-x)-p^{2}x(1-x)](\gamma_{\lambda}p_{\lambda'}+\gamma_{\lambda'}p_{\lambda})\right)\sigma_{\mu'\nu'}\psi(p)+\cdots=\nonumber\\
 &=&\frac{2g^{2}}{16\pi^{2}\epsilon}\kappa^{\mu\nu\lambda\rho}\kappa^{\mu'\nu'\lambda'\rho}\eta_{\rho\rho'}\sigma_{\mu\nu}\bar{\psi}(-p)\times\nonumber \\
 &\times& \left[
 (\frac{2m}{3}+\frac{\slashed{p}}{4})p_{\lambda}p_{\lambda}+\frac{1}{12}(\frac{p^2}{4}-m^2)(\gamma_{\lambda}p_{\lambda'}+\gamma_{\lambda'}p_{\lambda})-\right.
\nonumber\\&-&
\left.\frac{1}{4}\eta_{\lambda\lambda'}\left(
\frac{m^3}{2}+\frac{m^2\slashed{p}}{6}-\frac{mp^2}{6}-\frac{\slashed{p}p^2}{12}
%(\frac{m}{3}+\frac{\slashed{p}}{4})p^2-(\frac{m}{2}+\frac{\slashed{p}}{3})(m^2+p^2)+(m+\frac{\slashed{p}}{2})m^2
\right) 
 \right]\sigma_{\mu'\nu'}\psi(p)+\cdots,
\end{eqnarray}
where the ellipsis stand for finite terms which we do not quote here.

In summary, we calculated the one-loop LV contributions arising in
the two point vertex functions due to the presence of the non minimal
pseudotensor LV coupling in Eq.\,\eqref{eq:defModel}. All these
contributions are divergent, so that for consistency we are enforced
to assume that these structures already exist in the tree level photon
Lagrangian, so that one has enough counterterms available to absorb
these divergences. This renormalization procedure leave us with arbitrary
finite counterterms, which have to be fixed by some physical conditions,
or by comparison with the experiment. This scenario is not new: the
same happens when the tensor coupling (missing the $\gamma_{5}$ present
in our LV vertex) is used to generate the usual CPT even $k_{F}$
coefficient in the SME, as discussed in\,\cite{Casana3} (see also\,\cite{Belich1}).
A somewhat different sort of arbitrariness appears in the perturbative
generation of the CFJ term\,\cite{JackiwKostelecky} and also the
axion-photon coupling\,\cite{BorgesAxion}, where the quantum corrections
are finite but ambiguous (regularization dependent). On the other
hand, it has been shown that the perturbative generation of aether-like
LV terms yielding well-defined, finite quantum corrections, happens
in different models\,\cite{aether}. We stress that our results are
to be interpreted within the framework of the effective field theory
approach \cite{Georgi}, according to which the non-renormalizable
models represent themselves as a low-energy effective description
of some more fundamental theory. In this context, coupling constants
of non-renormalizable operators have negative mass dimension, arising
as a consequence of integrating out some heavy modes, and being proportional
to negative powers of some large mass scale. It is worth to point
out that originally, the SME itself has been introduced as a low-energy
effective description of the string theory \cite{KosSam}. Therefore,
the appearance of non-renormalizable couplings and of divergent terms
in our theory is very natural.

The end result is that, to cancel one loop divergences, the effective
Maxwell Lagrangian, taking into account the LV terms that are generated
by the fermion loop, should look like 
\begin{equation}
{\cal L}^{eff}=-\frac{1}{4}\left(F_{\mu\nu}F^{\mu\nu}+r_{1}mg\bar{F}_{\mu\nu}\tilde{F}^{\mu\nu}+r_{2}m^{2}g^{2}\bar{F}_{\mu\nu}\bar{F}^{\mu\nu}+r_{3}g^{2}\bar{F}_{\mu\nu}\Box\bar{F}^{\mu\nu}+r_{4}g^{2}\partial^{\mu}\bar{F}_{\mu\nu}\partial_{\lambda}\bar{F}^{\lambda\nu}\right),\label{actgen}
\end{equation}
where the $r_{i}$ are dimensionless renormalization constants. The
terms proportionals to $r_{1}$ and $r_{2}$ are minimal LV operators,
being first and second order in the LV tensor, respectively, and therefore
should relate to the $k_{F}$ term of the photon sector of the minimal
SME, while those proportional to $r_{3}$ and $r_{4}$ are higher
derivative, non-minimal terms. Since we are interested in investigating
the basic properties induced in the photon sector by these LV operators,
we will not pursue the task of fixing the exact values of $r_{i}$
in the following, instead they will be assumed of order one, in order
to establish rough order-of-magnitude constraints on $\kappa$, as
well as other interesting physical consequences.

We close this section by making some comments about the extension
of our results for finite temperature which we elaborate within the
framework proposed in \cite{GrigSem} and further applied in \cite{MarizFiniteTemp,MarizNascimento14}
and other papers. To justify the validity of this approach, we remind
that within Lorentz-breaking theories, there are two types of Lorentz
transformations, the observer ones, which transform both dynamical
fields and background coefficients, and the particle ones, which transform
only the dynamical fields, see discussion in \cite{Kostelecky}. As
a result, the Lagrangian is invariant with respect to observer Lorentz
transformations but not particle ones, and in within the observer
viewpoint we apply the usual finite temperature methodology. We proceed
with the basic momentum integral in Eq.\,(\ref{int}), which appears
both in the minimal and non-minimal contributions. For simplicity,
we restrict ourselves to the minimal terms proportional to $\bar{F}^{\mu\nu}\bar{F}_{\mu\nu}$
and $\bar{F}^{\mu\nu}\tilde{F}_{\mu\nu}$, and the external momentum
$p$ in the denominator are set to zero. Following the Matsubara formalism,
we carry out the Wick rotation and discretize the Euclidean $p_{0}$
variable according to $p_{0}=2\pi T(n+\frac{1}{2})$, for integer
$n$. As a result, we have the following finite temperature result for our basic integral (\ref{int}):
\begin{equation}
I=T\sum\limits _{n=-\infty}^{\infty}\int\frac{d^{3}\vec{k}}{(2\pi)^{3}}\frac{1}{\left(\vec{k}^{2}+m^{2}+4\pi^{2}T^{2}(n+\frac{1}{2})^{2}\right)^{2}}\thinspace,
\end{equation}
which, after integration yields 
\begin{equation}
I\left(a\right)=\frac{1}{16\pi^{2}}\sum\limits _{n=-\infty}^{\infty}\frac{1}{[a^{2}+(n+\frac{1}{2})^{2}]^{\frac{1}{2}+\epsilon}}\thinspace,
\end{equation}
where $a=\frac{m}{2\pi T}$, and we have introduced the parameter
$\epsilon\to0$ since at $\epsilon=0$ the sum diverges. Using the
well-known sum formula \cite{MarizFiniteTemp}, we obtain 
\begin{eqnarray}
\label{intres}
I(a)=\frac{1}{16\pi^{2}\epsilon}-\frac{1}{4\pi^2}\int_{a}^{\infty}\frac{dz}{\sqrt{z^{2}-a^{2}}}\frac{1}{e^{2\pi z}+1}.
\end{eqnarray}
We note that the pole part explicitly reproduces the zero-temperature
result as it should be. The finite part of $I(a)$ vanishes in the
zero temperature limit ($a\to\infty$). 

 It is interesting to study the high-temperature ($a\to 0$) behavior of this result as well. To avoid the singularity in the lower limit of the integral above, we use the following analytic continuation formula (see f.e. \cite{MarizFiniteTemp}):
\begin{eqnarray}
&&\int_a^{\infty}\frac{dz}{(z^2-a^2)^{\lambda}}{\rm Re} \frac{1}{e^{2\pi(z+ib)}-1}
=\nonumber\\
&=&-\frac{1}{2a^2}\frac{3-2\lambda}{1-\lambda}\int_a^{\infty}\frac{dz}{(z^2-a^2)^{\lambda-1}}{\rm Re}\frac{1}{e^{2\pi(z+ib)}-1}
-\nonumber\\
&-&\frac{1}{4a^2}\frac{1}{(2-\lambda)(1-\lambda)}\int_a^{\infty}\frac{dz}{(z^2-a^2)^{\lambda-2}}\frac{d^2}{dz^2}{\rm Re}\frac{1}{e^{2\pi(z+ib)}-1}.
\end{eqnarray}
Applying it for our case $\lambda=1/2$, $b=1/2$, and taking into account that the our integral in (\ref{intres}) is real, we represent this integral as:
\begin{eqnarray}
&&\int_a^{\infty}\frac{dz}{(z^2-a^2)^{1/2}}\frac{1}{e^{2\pi z}+1}=\nonumber\\
&=&-\frac{1}{a^2}\left[\frac{1}{2}\int_a^{\infty}dz(z^2-a^2)^{1/2}(1-\tanh\pi z)
+\frac{\pi^2}{3}\int_a^{\infty}dz(z^2-a^2)^{3/2} \frac{\tanh \pi z}{\cosh^2\pi z}\right].
\end{eqnarray}
The integrand in the parentheses in the r.h.s. displays no singularity at $a\to 0$. Hence, this integral in the limit $a\to 0$ can be written as
\begin{eqnarray}
&&\int_a^{\infty}\frac{dz}{(z^2-a^2)^{1/2}}\frac{1}{e^{2\pi z}+1}\simeq
%\nonumber\\&\simeq&
-\frac{C_0}{a^2}+\ldots,
\end{eqnarray}
where dots are for the subleading at $a\to 0$ terms, and 
$$
C_0=\frac{1}{2}\int_0^{\infty}dz z (1-\tanh \pi z)+%\nonumber\\&+&
\frac{\pi^2}{3}\int_0^{\infty}dz z^3 \frac{\tanh \pi z}{\cosh^2\pi z}
$$
is a finite constant. Thus, recovering the temperature dependence through using the explicit expression of $a$ we can write the following higher-dimensional asymptotic form for $I(a)$:
\begin{eqnarray}
I(a)|_{a\to 0}=\frac{1}{16\pi^{2}\epsilon}+\frac{C_0}{4\pi^2a^2}+\ldots=\frac{1}{16\pi^{2}\epsilon}+C_0\frac{T^2}{m^2}+\ldots.
\end{eqnarray}
This result grows quadratically with the temperature. Such a behavior is not unusual, it occurs, for example, for some contributions in \cite{MarizNascimento14}.

The final minimal correction
to the photon Lagrangian, in the finite temperature case, turns out
to be 
\begin{equation}
{\cal L}_{T}^{eff}=2\left(m^{2}g^{2}\bar{F}^{\mu\nu}\bar{F}_{\mu\nu}-meg\bar{F}^{\mu\nu}\tilde{F}_{\mu\nu}\right)I(a),
\end{equation}
so the essential structure of the LV corrections induced in the photon
two point function is preserved in the case of finite temperature.

\section{\label{sec:The-Induced-Minimal}The Induced Minimal LV Term}

Inspired by the results of the previous section, we now consider the
following effective Lagrangian 
\begin{equation}
{\cal L}_{\left(1\right)}^{eff}=-\frac{1}{4}\left(F_{\mu\nu}F^{\mu\nu}+c_{1}\kappa_{\mu\nu\alpha\beta}\epsilon^{\alpha\beta\rho\sigma}F^{\mu\nu}F_{\rho\sigma}+c_{2}\kappa_{\mu\nu\alpha\beta}\kappa^{\alpha\beta\rho\sigma}F^{\mu\nu}F_{\rho\sigma}\right)\thinspace,\label{eq:FirstOrder}
\end{equation}
where $c_{1}$ and $c_{2}$ are dimensionless constants. These corrections
amount to a CPT even $k_{F}$ term in the photon sector of the SME,
\begin{equation}
{\cal L}_{{\rm SME}}\supset-\frac{1}{4}k_{F}^{\alpha\beta\rho\sigma}F_{\alpha\beta}F_{\rho\sigma}\thinspace,
\end{equation}
where 
\begin{equation}
k_{F}^{\alpha\beta\rho\sigma}=c_{1}\kappa^{\alpha\beta\mu\nu}\epsilon_{\mu\nu}^{\hphantom{\mu\nu}\rho\sigma}+c_{2}\kappa^{\alpha\beta\mu\nu}\kappa_{\mu\nu}^{\hphantom{\mu\nu}\rho\sigma}\thinspace.\label{eq:kfMinimal}
\end{equation}

In order to write this Lagrangian in a way that allows for more physical
insight, we choose to decompose $\kappa^{\mu\nu\alpha\beta}$ into
three $3\times3$ matrices $\kappa_{a}$, $\kappa_{b}$, $\kappa_{c}$
according to\begin{subequations}\label{eq:kappa3D} 
\begin{align}
\left(\kappa_{a}\right)^{ij} & =\kappa^{0ilm}\epsilon^{jlm}\thinspace,\\
\left(\kappa_{b}\right)^{ij} & =\kappa^{0i0j}\thinspace,\\
\left(\kappa_{c}\right)^{ij} & =\epsilon^{ikm}\epsilon^{jpq}\kappa^{kmpq}\thinspace.
\end{align}
\end{subequations}This is the same kind of decomposition used for
the $k_{F}$ tensor in the photon sector of the SME\,\cite{LVQED},
but here applied to a different object, whose symmetry properties
are the same. Due to the symmetries of $\kappa^{\mu\nu\lambda\rho}$,
one can see that $\kappa_{b}$ and $\kappa_{c}$ are symmetric matrices
by definition.

Recalling the definition of the dual electromagnetic field tensor\,\eqref{eq:dualF},
together with the relations between the covariant tensors and the
vectorial fields ${\bf E}$ and ${\bf B}$, 
\begin{equation}
F^{i0}=E^{i},\thinspace F^{ij}=-\epsilon^{ijk}B^{k},\thinspace\tilde{F}^{i0}=-B^{i},\thinspace\tilde{F}^{ij}=-\epsilon^{ijk}E^{k}\thinspace,
\end{equation}
we can cast Eq.\,\eqref{eq:FirstOrder} as 
\begin{multline}
{\cal L}_{\left(1\right)}^{eff}=\frac{1}{2}\left({\bf E}^{2}-{\bf B}^{2}\right)+\frac{1}{2}{\bf E}\cdot\left[2c_{1}\kappa_{a}+c_{2}\left(4\kappa_{b}^{2}-\kappa_{a}\kappa_{a}^{T}\right)\right]\cdot{\bf E}\\
+{\bf E}\cdot\left[c_{1}\left(2\kappa_{b}-\frac{1}{2}\kappa_{c}\right)+\frac{1}{2}c_{2}\kappa_{a}\kappa_{c}-2c_{2}\kappa_{b}\kappa_{a}\right]\cdot{\bf B}\\
-\frac{1}{2}{\bf B}\cdot\left(2c_{1}\kappa_{a}-c_{2}\kappa_{a}^{T}\kappa_{a}+\frac{1}{4}c_{2}\kappa_{c}^{2}\right)\cdot{\bf B}\thinspace.\label{eq:FirstOrder3D}
\end{multline}
It is instructive to compare this with the similar terms present in
the SME, 
\begin{equation}
{\cal L}_{photon}^{SME}=\frac{1}{2}\left({\bf E}^{2}-{\bf B}^{2}\right)+\frac{1}{2}{\bf E}\cdot\kappa_{DE}\cdot{\bf E}+{\bf E}\cdot\kappa_{DB}\cdot{\bf B}-\frac{1}{2}{\bf B}\cdot\kappa_{HB}\cdot{\bf B}\thinspace.\label{eq:photonSME}
\end{equation}
We see that different combinations of $\kappa_{a}$, $\kappa_{b}$
and $\kappa_{c}$ contribute to the coefficients $\kappa_{DE}$, $\kappa_{DB}$
and $\kappa_{HB}$.

To simplify the analysis, we will first look at the first order terms
in\,\eqref{eq:FirstOrder3D}, which are naturally the dominant ones.
In this approximation, we see that the $\kappa_{a}$ coefficients
are responsible for generating $\kappa_{DE}$ and $\kappa_{HB}$ terms,
while the combination $2\kappa_{b}-\frac{1}{2}\kappa_{c}$ contributes
with a $\kappa_{DB}$ term, according to the correspondence\begin{subequations}\label{eq:minimal_correspondence}
\begin{align}
\kappa_{DE} & =\kappa_{HB}=2c_{1}\kappa_{a}\thinspace,\\
\kappa_{DB} & =c_{1}\left(2\kappa_{b}-\frac{1}{2}\kappa_{c}\right)\thinspace,
\end{align}
\end{subequations}where we used the fact that both $\kappa_{b}$
and $\kappa_{c}$ are symmetric matrices. We also recall that the
LV coefficients in Eq.\,\eqref{eq:photonSME} can be rewritten in
terms of CPT even and odd parts as follows\,\cite{LVQED},\begin{subequations}\label{eq:CPTeven_odd}
\begin{align}
\left(\kappa_{e+}\right) & =\frac{1}{2}\left(\kappa_{DE}+\kappa_{HB}\right)\thinspace,\thinspace\left(\kappa_{e-}\right)=\frac{1}{2}\left(\kappa_{DE}-\kappa_{HB}\right)-\frac{1}{3}\text{tr}\,\kappa_{DE}\thinspace,\\
\left(\kappa_{o\pm}\right) & =\frac{1}{2}\left(\kappa_{DB}\pm\kappa_{HE}\right)=\frac{1}{2}\left(\kappa_{DB}\mp\kappa_{DB}^{T}\right)\,,\\
\kappa_{\text{tr}} & =\frac{1}{3}\text{tr}\,\kappa_{DE}\thinspace,
\end{align}
\end{subequations}where $\kappa_{e}$ and $\kappa_{o}$ are CPT even
and odd coefficients, respectively. In the photon sector of the SME,
the CPT even coefficient $\kappa_{F}^{\mu\nu\rho\sigma}$ is assumed
to have vanishing double trace $\left(\kappa_{F}\right)_{\hphantom{\mu\nu}\mu\nu}^{\mu\nu}=0$,
since any non-vanishing value for $\left(\kappa_{F}\right)_{\hphantom{\mu\nu}\mu\nu}^{\mu\nu}$
could be reabsorbed in a normalization of the usual kinetic term;
this leads to the vanishing trace of the combination $\kappa_{DE}+\kappa_{HB}$,
which amounts to $\text{tr}\kappa_{DE}=-\text{tr}\kappa_{HB}$. For
the specific case of our model, in the currently considered approximation,
the correspondence given in Eq.\,\eqref{eq:minimal_correspondence}
means that any non-vanishing trace of $\kappa_{a}$ should lead to
no observable effect in the photon sector, and therefore we will assume
${\rm tr}\,\kappa_{a}=0$ for the moment.

Applying the decomposition given in Eq.\,\eqref{eq:CPTeven_odd}
for the effective LV coefficients generated in our model, according
to Eq.\,\eqref{eq:minimal_correspondence}, leads to 
\begin{align}
\left(\kappa_{e+}\right) & =2c_{1}\kappa_{a}\thinspace,\thinspace\left(\kappa_{e-}\right)=0\thinspace,\thinspace\kappa_{\text{tr}}=0\thinspace,\\
\left(\kappa_{o+}\right) & =0\thinspace,\thinspace\left(\kappa_{o-}\right)=c_{1}\left(2\kappa_{b}-\frac{1}{2}\kappa_{c}\right)\,.
\end{align}
Therefore, we have generated non-vanishing coefficients $\kappa_{e+}$
and $\kappa_{o-}$, which are responsible for birefringence effects
in the propagation of light in vacuum and, as a consequence, have
very strong experimental constraints: typical bounds are of order
$10^{-37}$ from astrophysical observations, and $10^{-15}$ from
laser interferometry\,\cite{DataTables}. We stress, however, that
translating these bounds on $\kappa_{e+}$ and $\kappa_{o-}$ to precise
bounds on $\kappa_{a}$ and $\kappa_{b}+\kappa_{c}$ should take into
account the mass of the integrated fermion, the coupling constant
$g$ and, most importantly, the renormalization constant $r_{1}$.
Assuming $r_{1}$ to be of order one, we can state the estimate 
\begin{equation}
mg\kappa_{a}<10^{-37}\thinspace,\thinspace\thinspace mg\left(2\kappa_{b}-\frac{1}{2}\kappa_{c}\right)<10^{-37}\thinspace,
\end{equation}
for the dimensionless combinations of fermion mass, coupling constant
and the LV parameters. For the electron mass of order $10^{-4}\,\text{GeV}$,
that amounts to constraints of order $10^{-33}\,\text{GeV}^{-1}$
to the corresponding coefficients in $H_{F}^{\left(5\right)\mu\nu\alpha\beta}$,
according to Eq.\,\eqref{eq:H5correspondence}, while for the proton
mass of order $1\,\text{GeV},$ the bounds would of order $10^{-37\,}\text{GeV}^{-1}$.

Now taking into account the second order terms, the correspondence
\eqref{eq:minimal_correspondence} changes to\begin{subequations}
\begin{align}
\kappa_{DE} & =2c_{1}\kappa_{a}+c_{2}\left(4\kappa_{b}^{2}-\kappa_{a}\kappa_{a}^{T}\right)\thinspace,\\
\kappa_{DB} & =c_{1}\left(2\kappa_{b}-\frac{1}{2}\kappa_{c}\right)+\frac{c_{2}}{2}\kappa_{a}\kappa_{c}-2c_{2}\kappa_{b}\kappa_{a}\thinspace,\\
\kappa_{HB} & =2c_{1}\kappa_{a}-c_{2}\kappa_{a}^{T}\kappa_{a}+\frac{c_{2}}{4}\kappa_{c}^{2}\thinspace,
\end{align}
\end{subequations}The new aspects arising from the second order contributions
are that now the condition of zero trace of $\kappa_{a}$ is not enough
to ensure that $\kappa_{DE}+\kappa_{HB}$ is traceless, and in general
we will have $\kappa_{tr}\neq0$. Also, the non-birefringent coefficients
$\kappa_{e-}$ and $\kappa_{o+}$ acquire non-vanishing values, which
are second order in $\kappa_{a}$, $\kappa_{b}$ and $\kappa_{c}$.
Typical experimental constraints for these non-birefringent coefficients
are of order $10^{-18}$ from astrophysics, and $10^{-15}$ from laser
interferometry\,\cite{DataTables}, but from these we will not try
to infer new constraints on the LV coefficients since they apply to
second-order combinations of $\kappa_{a}$, $\kappa_{b}$ and $\kappa_{c}$,
and therefore could provide at the best very modest constraints.

\section{Covariant Dispersion Relations: the minimal case}

To fully unveil the birefringence effects resulting from the minimal
LV model defined in Eq.\,(\ref{eq:FirstOrder}), we calculate the
dispersion relations using the formalism presented in\,\cite{nonMinimalLV1}.
The general idea is to use a plane wave ansatz $A_{\mu}\left(x\right)=A_{\mu}\left(p\right)e^{-ip\cdot x}$
and write the covariant equations of motion for the electromagnetic
potential in the absence of sources in the form 
\begin{equation}
M^{\mu\nu}\left(p\right)A_{\mu}\left(p\right)=0\thinspace.\label{eq:coveom}
\end{equation}
For a LV Lagrangian of the general form we will be interested in,
\begin{equation}
{\cal L}=-\frac{1}{4}F^{\mu\nu}F_{\mu\nu}-\frac{1}{4}F_{\mu\nu}\left(\hat{\kappa}_{F}\right)^{\mu\nu\alpha\beta}F_{\alpha\beta}\thinspace,
\end{equation}
the matrix $M$ assumes the explicit form 
\begin{equation}
M^{\mu\nu}=2\hat{\chi}^{\mu\alpha\nu\beta}p_{\alpha}p_{\beta}\thinspace,
\end{equation}
where 
\begin{equation}
\hat{\chi}^{\mu\alpha\nu\beta}=\frac{1}{2}\left(\eta^{\mu\nu}\eta^{\alpha\beta}-\eta^{\mu\alpha}\eta^{\nu\beta}\right)+\left(\hat{\kappa}_{F}\right)^{\mu\alpha\nu\beta}\thinspace.\label{eq:defChi}
\end{equation}
Gauge symmetry implies that $M^{\mu\nu}p_{\nu}=0$, so that Eq.\,\eqref{eq:coveom}
always have the trivial, pure gauge solution $A_{\mu}\sim p_{\mu}$.
This leads to the conclusion that $\det M=0$, so $M$ has null spaces,
i.e., its rank is smaller than its dimension. By carefully studying
the null space structure of $M$, using exterior algebra tools, an
explicit, covariant form for the dispersion relation can be shown
to be\,\cite{nonMinimalLV1} 
\begin{equation}
\epsilon_{\mu_{1}\mu_{2}\mu_{3}\mu_{4}}\epsilon_{\nu_{1}\nu_{2}\nu_{3}\nu_{4}}p_{\rho_{1}}p_{\rho_{2}}p_{\rho_{3}}p_{\rho_{4}}\hat{\chi}^{\mu_{1}\mu_{2}\nu_{1}\rho_{1}}\hat{\chi}^{\nu_{2}\rho_{2}\rho_{3}\mu_{3}}\hat{\chi}^{\rho_{4}\mu_{4}\nu_{3}\nu_{4}}=0\thinspace.\label{eq:covariantDR}
\end{equation}
This equation can be applied for our model, after the proper identification
of the coefficient $\left(\hat{\kappa}_{F}\right)^{\mu\alpha\nu\beta}$
generated in the photon sector by the radiative corrections, which,
in the minimal case, is given by Eq.\,\eqref{eq:kfMinimal}.

We will perform the calculation of the dispersion relation for a particular
case. Taking into account the decomposition \eqref{eq:kappa3D}, we
set $\kappa_{b}=\kappa_{c}=0$, and take $\kappa_{a}$ as the antisymmetric
matrix 
\begin{equation}
\kappa_{a}=\left(\begin{array}{ccc}
0 & \kappa^{3} & -\kappa^{2}\\
-\kappa^{3} & 0 & \kappa^{1}\\
\kappa^{2} & -\kappa^{1} & 0
\end{array}\right)\thinspace,\label{eq:kappaparticular}
\end{equation}
so that the LV is parametrized by a vector ${\bf k}=\left(\kappa^{1},\kappa^{2},\kappa^{3}\right)$.
By expanding the expression\,\eqref{eq:covariantDR} with this particular
choice of $\kappa$, we find a second order polynomial equation in
$\left(p_{0}\right)^{2}$, which can be solved to find 
\begin{equation}
\left(p_{0}\right)^{2}=\frac{\Theta\pm\sqrt{\Xi}}{6\left(c_{2}{\bf k}^{2}-1\right){}^{2}-8c_{1}^{2}{\bf k}^{2}}\thinspace,\label{eq:DRminimal}
\end{equation}
where 
\begin{multline}
\Theta=6{\bf p}^{2}+{\bf k}^{2}{\bf p}^{2}\left(6c_{2}^{2}{\bf k}^{2}+8c_{1}^{2}-12c_{2}\right)\\
+\left({\bf p}\times{\bf k}\right)^{2}\left(-12c_{2}^{3}\left({\bf k}^{2}\right)^{2}+12c_{2}^{2}{\bf k}^{2}+16c_{1}^{2}c_{2}{\bf k}^{2}-48c_{1}^{2}\right)\thinspace,
\end{multline}
and 
\begin{multline}
\Xi=\left\{ 3\left(1-c_{2}{\bf k}^{2}\right)\left[2{\bf p}^{2}+4c_{2}^{2}{\bf k}^{2}\left({\bf p}\times{\bf k}\right)^{2}-2c_{2}{\bf k}^{2}{\bf p}^{2}\right]\right.\\
\left.+4c_{1}^{2}\left[4\left({\bf p}\times{\bf k}\right)^{2}\left(c_{2}{\bf k}^{2}-3\right)+2{\bf k}^{2}{\bf p}^{2}\right]\right\} ^{2}\\
+4\left(3\left(c_{2}{\bf k}^{2}-1\right){}^{2}-4c_{1}^{2}{\bf k}^{2}\right)\left(4c_{2}\left({\bf p}\times{\bf k}\right)^{2}-{\bf p}^{2}\right)\\
\times\left[3\left(c_{2}{\bf k}^{2}-1\right)\left(c_{2}\left({\bf k}\cdot{\bf p}\right)^{2}-{\bf p}^{2}\right)-4c_{1}^{2}\left({\bf k}\cdot{\bf p}\right)^{2}\right]\thinspace.
\end{multline}
One may verify that the usual result $\left(p_{0}\right)^{2}=\left({\bf p}\right)^{2}$
is obtained when $\kappa\rightarrow0$. In the general case in which
$\Xi\neq0$, equation\,\eqref{eq:DRminimal} exhibits the expected
birefringence in the model, as predicted in the previous section.
The propagation of light signals in this theory is also anisotropic,
since the phase velocity contains terms which depend on the relative
orientation of ${\bf p}$ and ${\bf k}$.

One interesting question that can be addressed with this result is
the causality of the wave propagation. We can obtain from Eq.\,\eqref{eq:DRminimal}
the phase, group, and front velocity for electromagnetic waves, by
means of 
\begin{equation}
v_{\text{phase}}=\frac{p^{0}}{|{\bf p}|},\thinspace v_{\text{group}}=\frac{dp^{0}}{d|{\bf p}|},\thinspace v_{\text{front}}=\lim_{|{\bf p}|\rightarrow\infty}v_{\text{phase}}\thinspace,
\end{equation}
and we say causality is ensured at the classical level if $v_{\text{group}}\leq1$
and $v_{\text{front}}\leq1$\,\cite{Sexl}. We consider the dispersion
relation for two particular cases, namely, waves propagating in the
same direction as ${\bf k}$, as well as in a perpendicular direction.
To simplify the resulting expressions, we consider that the LV parameters
are very small, and so expand the results up to the second order in
$|{\bf k}|$. We also verify that, in our case, the equality $v_{\text{phase}}=v_{\text{group}}=v_{\text{front}}$
holds for every case.

For ${\bf p}$ parallel to ${\bf k}$, we choose, without loss of
generality, ${\bf p}=\left(p,0,0\right)$ and ${\bf k}=\left(\kappa,0,0\right)$,
and we obtain, 
\begin{equation}
v_{\text{group}}=1\pm\frac{2}{\sqrt{3}}\kappa\left|c_{1}\right|+\frac{2}{3}\kappa^{2}c_{1}^{2}+{\cal O}\left(\kappa^{3}\right)\quad\left(\text{for }{\bf p}\parallel{\bf k}\right)\thinspace,
\end{equation}
as for the case of ${\bf p}$ perpendicular to ${\bf k}$, we choose
${\bf p}=\left(p,0,0\right)$ and ${\bf k}=\left(0,0,\kappa\right)$,
thus obtaining 
\begin{equation}
v_{\text{group}}=1+\frac{1}{6}\kappa^{2}\left(2c_{1}^{2}\pm\left|2c_{1}^{2}+3c_{2}\right|\right)+{\cal O}\left(\kappa^{3}\right)\quad\left(\text{for }{\bf p}\perp{\bf k}\right)\thinspace.
\end{equation}
We conclude that for the parallel case, one of the polarizations is
generally non-causal, while for the perpendicular case, if $c_{2}<0$,
we can ensure that both polarizations are causal, otherwise one of
them will violate causality.

\section{The induced non minimal LV sector}

We now study the non minimal LV terms induced in the photon sector,
corresponding to the ones proportional to $r_{3}$ and $r_{4}$ in
Eq.\,\eqref{actgen}: both contribute to the non minimal $\hat{k}_{F}$
term in the SME\,\cite{nonMinimalLV1}, 
\begin{equation}
{\cal L}_{{\rm SME}}\supset-\frac{1}{4}\hat{k}_{F}^{\alpha\beta\rho\sigma}F_{\alpha\beta}F_{\rho\sigma}\thinspace.
\end{equation}
These terms are expected to be subdominant, so we will provide a more
simplified discussion, in order to provide insight into the kind of
effects that could be generated at this level, yet pointing out that
a more complete discussion, including for example higher loop orders,
would be necessary in order to provide conclusive results.

For simplicity, we will consider the two non-minimal terms separately,
starting with 
\begin{equation}
{\cal L}_{\left(2\right)}^{eff}=-\frac{1}{4}\left(F_{\mu\nu}F^{\mu\nu}+c_{3}\kappa_{\mu\nu\rho\sigma}\kappa^{\mu\nu\alpha\beta}F^{\rho\sigma}\Box F_{\alpha\beta}\right)\thinspace.
\end{equation}
With the identification 
\begin{equation}
\hat{k}_{F}^{\alpha\beta\rho\sigma}=-c_{3}\kappa^{\mu\nu\rho\sigma}\kappa_{\mu\nu}^{\hphantom{\mu\nu}\alpha\beta}p^{2}\thinspace,
\end{equation}
where the $\kappa^{\mu\nu\rho\sigma}$ tensor is decomposed according
to Eq.\,\eqref{eq:kappa3D}, and using the standard definitions for
the non minimal coefficients $\hat{\kappa}_{DE}$, $\hat{\kappa}_{DB}$
and $\hat{\kappa}_{HB}$, \begin{subequations} \label{eq:defKappa}
\begin{align}
\left(\hat{\kappa}_{DB}\right)^{jk} & =\hat{\kappa}_{F}^{0jlm}\epsilon^{klm}\thinspace,\\
\left(\hat{\kappa}_{DE}\right)^{jk} & =-2\hat{\kappa}_{F}^{0j0k}\thinspace,\\
\left(\hat{\kappa}_{HB}\right)^{jk} & =\frac{1}{2}\epsilon^{jrm}\epsilon^{kpq}\hat{\kappa}_{F}^{rmpq}\thinspace.
\end{align}
\end{subequations}one obtains directly\begin{subequations} 
\begin{align}
\left(\hat{\kappa}_{DE}\right)^{jk} & =-4c_{3}^{2}\left(\kappa_{b}^{2}-\frac{1}{4}\kappa_{a}\kappa_{a}^{T}\right)^{jk}p^{2}\thinspace,\\
\left(\hat{\kappa}_{HB}\right)^{jk} & =c_{3}^{2}\left(\kappa_{a}^{T}\kappa_{a}-\frac{1}{4}\kappa_{c}^{2}\right)^{jk}p^{2}\thinspace,\\
\left(\hat{\kappa}_{DB}\right)^{jk} & =2c_{3}^{2}\left(\kappa_{a}^{T}\kappa_{b}-\frac{1}{4}\kappa_{a}\kappa_{c}\right)^{jk}p^{2}\thinspace.
\end{align}
\end{subequations} From this result, one can use the non minimal
generalization of Eq.\,\eqref{eq:CPTeven_odd} to calculate the coefficients
$\hat{\kappa}_{e\pm}$, $\hat{\kappa}_{o\pm}$ and $\hat{\kappa}_{\text{tr}\pm}$\,\cite{nonMinimalLV1}.
Instead of quoting the exact expressions, we comment on their general
features. First, the trace components $\hat{\kappa}_{\text{tr}\pm}$
are generically nonzero, in the sense there is not a simple condition
on $\kappa_{a}$, $\kappa_{b}$ or $\kappa_{c}$ that can ensure $\hat{\kappa}_{\text{tr}\pm}=0$,
as in the first order, minimal term analyzed in Sec.\,\ref{sec:The-Induced-Minimal}.
Second, we notice that $\hat{\kappa}_{e+}\sim\hat{\kappa}_{DE}+\hat{\kappa}_{HB}$
is nonzero whenever any of the $\kappa_{a}$, $\kappa_{b}$ or $\kappa_{c}$
is nonzero (except for very specific values in which their contribution
to $\hat{\kappa}_{e+}$ cancel), and since the $\hat{\kappa}_{e+}$
is associated with birefringence, we can state that birefringence
is a generic feature of this model. The other birefringent coefficient,
$\hat{\kappa}_{o-}$, can only be nonzero if both $\kappa_{a}$ and
$\kappa_{c}$, or $\kappa_{a}$ and $\kappa_{b}$, are nonzero.

Next, we consider the remaining non minimal term, as in 
\begin{equation}
{\cal L}_{\left(2\right)}^{eff}=-\frac{1}{4}\left(F_{\mu\nu}F^{\mu\nu}+c_{4}\kappa_{\mu\nu\rho\sigma}\kappa^{\lambda\nu\alpha\beta}\partial^{\mu}F^{\rho\sigma}\partial_{\lambda}F_{\alpha\beta}\right)\thinspace,\label{eq:L4}
\end{equation}
and the identification with the SME $\hat{k}_{F}^{\alpha\beta\rho\sigma}$
coefficient now reads 
\begin{equation}
\hat{k}_{F}^{\alpha\beta\rho\sigma}=-c_{4}\kappa^{\mu\nu\rho\sigma}\kappa_{\lambda\nu}^{\hphantom{\mu\nu}\alpha\beta}p_{\mu}p^{\lambda}\thinspace.
\end{equation}
The calculation of the corresponding $\hat{\kappa}_{DE}$, $\hat{\kappa}_{DB}$
and $\hat{\kappa}_{HB}$ is more involved in this case, but can also
be carried out directly. We will not quote the cumbersome expressions
that result, but we comment that birefringence is also a generic consequence
of the Lagrangian in Eq.\,\eqref{eq:L4}.

We can gain more insight into wave propagation in this model with
the help of the covariant dispersion given in Eq.\,\eqref{eq:covariantDR},
where now, we use 
\begin{equation}
\hat{\kappa}_{F}^{\rho\sigma\alpha\beta}=-\left(c_{3}\kappa^{\mu\nu\rho\sigma}\kappa_{\mu\nu}^{\hphantom{\mu\nu}\alpha\beta}p^{2}+c_{4}\kappa^{\mu\nu\rho\sigma}\kappa_{\lambda\nu}^{\hphantom{\mu\nu}\alpha\beta}p_{\mu}p^{\lambda}\right)\thinspace.
\end{equation}
As before, we select a particular case to show what kind of physical
effects we can expect in this model. Again we consider $\kappa_{b}=\kappa_{c}=0$
and $\kappa_{a}$ given as in Eq.\,\eqref{eq:kappaparticular}. Expanding
the dispersion relation in Eq.\,\eqref{eq:covariantDR} we obtain
an eighth order polynomial in $p^{0}$, whose solutions can be found
in explicit form with a computer algebra system such as \emph{Mathematica}.
The independent solutions are\begin{subequations} 
\begin{align}
\left(p^{0}\right)_{a}^{2} & ={\bf p}^{2}\thinspace,\\
\left(p^{0}\right)_{b}^{2} & =\frac{1+c_{3}\left({\bf k}\cdot{\bf p}\right)^{2}}{c_{3}{\bf k}^{2}}\thinspace,\\
\left(p^{0}\right)_{c}^{2} & =\frac{1+c_{3}{\bf k}^{2}{\bf p}^{2}+2c_{4}\left({\bf p}\times{\bf k}\right)^{2}}{c_{3}{\bf k}^{2}}\thinspace.
\end{align}
\end{subequations} We find therefore three modes for electromagnetic
wave propagation, one being completely independent of the LV background,
corresponding to the usual wave propagation in Maxwell electrodynamics.
The other two modes are inherently Lorentz violating, in the sense
they do not possess a smooth limit when ${\bf k}\rightarrow{\bf 0}$,
the corresponding poles in the complex $p^{0}$ plane going to infinity
in this limit.

For the first LV mode, $\left(p^{0}\right)_{b}$, we calculate the
phase, group and front velocity, for the parallel case (i.e., ${\bf p}=\left(p,0,0\right)$
and ${\bf k}=\left(\kappa,0,0\right)$) as well as for the perpendicular
case (i.e. ${\bf p}=\left(p,0,0\right)$ and ${\bf k}=\left(0,0,\kappa\right)$.
The results are 
\begin{equation}
v_{\text{phase}}=\sqrt{1+\frac{1}{c_{3}p^{2}\kappa^{2}}},\thinspace v_{\text{group}}=\left(v_{\text{phase}}\right)^{-1},\thinspace v_{\text{front}}=1\quad\left(\text{for }{\bf p}\parallel{\bf k}\right)\thinspace,
\end{equation}
and 
\begin{equation}
v_{\text{phase}}=\frac{1}{\sqrt{c_{3}p^{2}\kappa^{2}}}=v_{\text{front}},\thinspace v_{\text{group}}=0\quad\left(\text{for }{\bf p}\perp{\bf k}\right)\thinspace.
\end{equation}
Despite $\left(p^{0}\right)_{b}$ propagating with $v_{\text{phase}}>1$
in the parallel direction to ${\bf k}$, this mode does not violate
causality because both $v_{\text{group}}$ and $v_{\text{front}}$
are less or equal to one. In the perpendicular direction, $\left(p^{0}\right)_{b}$
is actually independent of ${\bf p}$, so even if classically we can
say causality is preserved since $v_{\text{phase}}=v_{\text{front}}<1$,
it is hard to imagine that a consistent quantum interpretation can
be made for this mode. As for the second LV mode, $\left(p^{0}\right)_{c}$,
proceeding as before we obtain the same results as for $\left(p^{0}\right)_{c}$
in the case ${\bf p}\parallel{\bf k}$. However, for ${\bf p}\perp{\bf k}$
we have 
\begin{align}
v_{\text{phase}} & =\sqrt{1+2\frac{c_{4}}{c_{3}}+\frac{1}{c_{3}p^{2}\kappa^{2}}},\thinspace v_{\text{front}}=\sqrt{1+2\frac{c_{4}}{c_{3}}},\\
v_{\text{group}} & =\sqrt{\frac{1+2\frac{c_{4}}{c_{3}}}{1+\kappa^{2}}}\quad\left(\text{for }{\bf p}\perp{\bf k}\right)\thinspace,
\end{align}
corresponding to non-causal wave propagation. We conclude that the
non-minimal piece of the Lorentz violating model has unphysical modes,
which is a general feature of the non-minimal SME extensions\,\cite{nonMinimalLV1,nonMInimalLV2}
(see for example\,\cite{Ferrnew} for a detailed discussion of dimension
six operators). From the phenomenological viewpoint, understanding
these models as effective field theories, one can say that these unphysical
modes are not expected to appear in low energy experiments, but a
deeper theoretical investigation about them is a non trivial and interesting
problem.

\section{Conclusions and Perspectives}

We considered the quantum impacts of a dimension-five pseudotensor
Lorentz-breaking spinor-vector coupling, by calculating the one loop
contributions to the two-point function of the gauge field. These
corrections turned out to be divergent, thus requiring the introduction
of the corresponding counterterms in the purely gauge sector in order
to eliminate these divergences. Therefore, the problem of studying
the new extended LV Maxwell theory naturally arose. For this theory,
we obtained the dispersion relations and found different modes of
wave propagation, with only some of them being consistent with regard
to the causality requirement. We found also that the birefringence
is a general feature in our resulting model, and from it some constraints
on the LV parameter could in principle be imposed.

The mechanism presented in this, as well as in other works in the
literature\,\cite{JackiwKostelecky,aether,Casana3,BorgesHDQED,BorgesAxion,Celeste},
involving Lorentz violating terms in the photon sector of the SME
arising as perturbative corrections originated from Lorentz violating
couplings in other sectors, could lead to the translation of the very
stringent bounds found in the photon sector to these original couplings.
Unfortunately, this perturbative corrections rarely appear without
some degree of uncertainty, due to ambiguities in the calculation
of Feynman diagrams, or to the renormalization procedure itself, as
we discussed in this work. This is certainly a subject that deserves
further study.

Further continuation of our study could also consist in a deeper discussion
of the impacts of the higher-derivative terms generated by the pseudotensor
coupling. The most interesting issue would be the investigation of
their influence on the unitarity of the theory. This question was
discussed for other non minimal couplings such as in\,\cite{Casana1},
and more recently unitarity in the presence of a Lorentz violating
three-derivative term appeared in\,\cite{CMR,CMR2}, so, performing
a similar analysis for the four-derivative term would be an interesting
problem.

\bigskip{}

\textbf{Acknowledgments.} The authors would like to thank V. A. Kostelecky
for discussions and interesting insights that helped us to improve
our paper. This work was partially supported by Conselho Nacional
de Desenvolvimento Científico e Tecnológico (CNPq) and Fundação de
Amparo à Pesquisa do Estado de São Paulo (FAPESP), via the following
grants: CNPq 304134/2017-1 and FAPESP 2017/13767-9 (AFF), CNPq 303783/2015-0
(AYP).

\end{document}